\documentclass[aps,prl,twocolumn,showpacs]{revtex4}
\usepackage{latexsym}
\usepackage{epsfig}
\usepackage{bm}
\usepackage{dcolumn}

\begin{document}

 \title{ Simulations on the Heavy Hadron Transport at RHIC }

 \author{Hong Miao$^1$, Chongshou Gao$^{2}$}
 \affiliation{$^1$Physics Department, Tsinghua University, Beijing
100084, China} \affiliation{$^2$School of Physics, Peking
University, Beijing 100871, China}

\begin{abstract}

Based on the hadron transport frames, detailed simulations are
carried out to discuss $\phi$, $\Omega$ productions and the
significant enhancements in the very low $p_T$ region for some of
the soft spectra in RHIC. Elastic interactions are introduced in
the simulations. The elastic cross sections vary from different
hadrons and energy scales, which can qualitatively explain the
different collective motions of various hadrons.


\end{abstract}





\maketitle

\section {Introduction} %
\label{sec:intro}

In the recent progress \cite{sign:transMod}, we have developed a
transport model to describe the hadron production in relativistic
heavy ion collisions. In the model, a decoupling hypersurface of
the fluid which consists of a group of splitting QGP droplets
\cite{sign:HBT_droplets} is used to emit hadrons. The model
satisfies most of light hadrons' spectra, but disagrees with the
distributions of $\phi$ \cite{sign:exp_STAR_phi,
sign:exp_PHENIX_phi}, $\Omega$ \cite{sign:exp_STAR_Strange} and
some heavy resonances \cite{sign:exp_STAR_Sig1385,
sign:exp_STAR_Xi1530}. As another remarkable phenomenon, there
seems to be significant enhancements at very low transverse
momentum in some of the spectra for $\Lambda$
\cite{sign:exp_STAR_Strange2}, $\Sigma^*(1385)$ and $\phi$, to
change the slopes. On the contrary, the model predicts a decrease
when $p_T$ drops. 

The idea of the hypersurface hadronization locks the emitted
hadrons in the local rest frames of the hypersurface. This is
right when the fluid is dense enough in the early stage of the
evolution. In the later stage, when the inelastic free path
becomes as long as $4\;fm$, the emission fixed in the hypersurface
might be less strict. Considering the spectra of light hadrons are
well satisfied, as other hydrodynamic methods with decoupling
hypersurface did, there must be some reasons to push the inside
light hadrons to participate in the collective motions. One of the
possible candidate is the elastic collision
\cite{sign:ZhuPartonwind}. Although elastic collisions do not
change the hadron abundances themselves, the effect of preventing
hadrons from entering the more dense regions and being absorbed,
will enhance their abundances or momenta.

Thus, detailed simulations on hadron transport are applied to some
of the heavy hadrons, such as $\phi$ and $\Omega$ in the central
events in RHIC Au-Au 200 $GeV$. The advantages for $\phi$ and
$\Omega$ are based on the condition that almost no hadrons could
decay to them and their widths are large enough to neglect the
affections of decays in the expanding medium. As a simplification,
only selected hadrons for the measurements are simulated. The
medium evolution and the feedbacks to the medium are not simulated
in microscopic frames. As the cross sections of some heavy or
multi-strange hadrons may be much smaller than those of light
hadrons \cite{sign:phi_cross}, it will provide us a possible
solution to explain the various collective behaviors between
different hadrons and $p_T$ regions.

\section {Medium Evolution} %
\label{sec:medium}

The decoupling of the fluid is a gradual progress in principle. As
the feedback of the hadron simulation to the fluid is unjustified,
the medium is separated to an ideal hydrodynamic part and a free
expanding part with evaporation by a cut $\epsilon_b$. The $1+1$
hydrodynamic evolution with an invariant boost for the central
events in RHIC Au-Au 200 $GeV$ is determined by the method in ref
\cite{sign:transMod}. The decoupling surface here is similar to
the results of surface evaporation \cite{sign:transMod} except in
the very early stage which is considered to be less important. For
a group of decoupled droplets, there are no collective motions and
local rest frames indeed. Approximately, the whole evolution is
defined,
\begin{equation}
 \frac{\partial{\epsilon}}{\partial{t}} =
    \int\{\epsilon_c V_R \frac{\partial{n_{vR}}}{\partial{t}}
    +\epsilon_c{n_{vR}}\frac{\partial{V_{R}}}{\partial{t}}\}dRdv,
 \label{eq:decouple_evo}
\end{equation}
where the former term in the integral stands for the free flights
of the droplets and the latter term is the evaporation. Then,
\begin{equation}
 \frac{\partial{n_{vR}}}{\partial{t}} =
 -v\frac{\partial{n_{vR}}}{\partial{r}}-(N-1)\frac{n_{vR}v}{r},
 \label{eq:decouple_density}
\end{equation}
where $n_{vR}({r},{v})=\frac{d^2{n}}{d{{v}}d{R}}$ is the velocity
and radius distribution of the number density of the droplets.

As we do not know the distribution of the droplets $n_{vR}$ and
can not simulate all of the droplets either, the expansion is
simplified as,
\begin{eqnarray}
 \frac{\partial{\lambda}}{\partial{t}} &=&
    -v\frac{\partial{\lambda}}{\partial{r}}-\lambda\frac{\partial{v}}{\partial{r}}-(N-1)\frac{\lambda{v}}{r},\\
 \frac{\partial{v}}{\partial{t}} &=&
    -v\frac{\partial{v}}{\partial{r}},
 \label{eq:decouple_expansion}
\end{eqnarray}
where $\lambda=\gamma\epsilon$ and $N=2$ is the symmetric
dimension. Considering that the droplet evolution by emitting and
absorbing hadrons is not clear, the evaporation term is omitted in
our calculations.

\section {Hadron Transport} %
\label{sec:transport}
The classical transport equation for the evolution of hadron phase
space distribution $f(t,{\bf r},{\bf p})$ can be written as
\begin{equation}
 \frac{\partial f}{\partial t} + {\bf v}\cdot\nabla f =
 Q + \alpha - \beta f,
\end{equation}
where the second term on the left hand side is the free streaming
part, and on the right hand side the lose and gain terms
$\beta({\bf p}) f$ and $\alpha({\bf p})$ indicate hadron
absorption and production in the medium, and $Q({\bf p};f)$ stands
for the elastic term.

In the simulations, hadrons generated in QGP or decayed from
heavier hadrons, will pass through the medium and decay at the
same time. Thus, the gain term includes two parts,
$\alpha=\alpha_{QGP}+\alpha_{dc}$. The decay contribution is easy
to obtain, but the term $\alpha_{QGP}$ is complicated. The reason
is that the quantitative temperature dependencies for some
parameters are unknown \cite{sign:VollEff}, especially in a
rapidly expanding system. Therefore, the production is simplified
as ideal QGP with massless quarks and gluons plus the
contributions of bound states which will never change via $T$, as
well as the coupling constants. The simplification is based on
such an idea that in large $T$ regions, where the inelastic free
path is short and the produced hadrons may travel more distance
before they escape from the medium, the affections may be less
important for the simulations. Thus, the approximation will be
sufficient enough if the productions are well satisfied near the
critical temperature $T_c$. Then,
\begin{equation}
 \alpha_{QGP} =\left\{
     \begin{array}{cc}
        \alpha_{0}\;(\frac{\epsilon'}{\epsilon_c'})^{\frac{3}{4}l}, &\quad\epsilon > \epsilon_c,\\
        \alpha_{0}\; \eta, &\quad\epsilon \leq \epsilon_c,
     \end{array}
 \right.
 \label{eq:pr}
\end{equation}
where $l=2$ for mesons and $l=1$ for baryons due to the
quark-diquark frame \cite{sign:diquark_miao}. $\eta$ is the ratio
of the occupation volume of all the QGP droplets to the whole
volume of the system, $\alpha_{0}$ is the production rate at $T_c$
and
\begin{eqnarray*}
 \epsilon'  =\epsilon-\epsilon_{Bound},\\
 \epsilon_c'=\epsilon_c-\epsilon_{Bound},
\end{eqnarray*}
where $\epsilon_{Bound}$ is contributions of bound states in QGP.

The loss rate includes the decay rate and the medium absorption,
$\beta=\beta_{dc}+\beta_{abs}$, where
$\beta_{dc}=\gamma^{-1}\Gamma$ and
\begin{equation}
 \beta_{abs} =\left\{
     \begin{array}{cc}
        \beta_{0}\;(\frac{\epsilon'}{\epsilon_c'})^{\frac{3}{4}}, &\quad\epsilon > \epsilon_c,\\
        \beta_{0}\; \eta, &\quad\epsilon \leq \epsilon_c,
     \end{array}
 \right.
 \label{eq:abs}
\end{equation}
with $\beta_{0}\approx
(1+v^2/3)n_q\langle{\sigma}\rangle_{inelastic}$ at the critical
temperature $T_c$. As the expression belongs in the local rest
frame of the medium, a Lorentz transformation on time step is
required when the simulation is applied in the selected frame.
Details on $\alpha_{0}$ and $\beta_{0}$ are discussed in ref
\cite{sign:transMod, sign:diquark_miao}.

It is noted, (\ref{eq:pr}) and (\ref{eq:abs}) are not independent,
even the droplets are assumed under random distribution. When
$\eta<1$, the location of production and absorption is correlated.
If the QGP droplets are large enough, a factor $\kappa(R)$ should
be inserted to the independent productions (\ref{eq:pr}) to remove
the absorption by the droplet where the hadron is produced by QGP
combination in principle. $\kappa(R)$ can be fitted as,
\begin{equation}
 \kappa(R)\sim \frac{1}{1+a\Delta+b\Delta^2},
\end{equation}
where
\begin{equation}
 \Delta \sim (1-\eta)\frac{\beta_0R}{v+v_s}.
\end{equation}
As discussed in above sections, the evolutions of the droplets are
unjustified so far. Thus we set $R \sim 0$ and $\kappa(R) \sim 1$
to minimize the parameter dependency.

Hadrons produced in time, position and momentum bins ($i$) are
tracked when they pass through the medium. They are recorded
separately, as
\begin{equation}
 N_i = N_{escape}+N_{reco},
\end{equation}
where $N_{escape}$ is the direct and reconstructed hadrons escaped
from the medium by cuts and $N_{reco}$ is the reconstructed
hadrons which may decay in the medium,
\begin{equation}
 N_{reco} = \sum_{step}\sum_I{\{N_I\prod_{j}{\frac{n_j}{N_I}}\}},
\end{equation}
with recorded decay products $n_j$ and
\begin{equation}
 N_I = Decayed \; hadrons \times Branch \; ratio.
\end{equation}

There is a small difference between our calculation and the
experimental manipulations. In the experiments, usually one
channel is measured for the reconstruction to provide the total
abundance by dividing the corresponding branch ratio.
\begin{figure}[h]
 \centering
 \includegraphics[width=.5\textwidth]{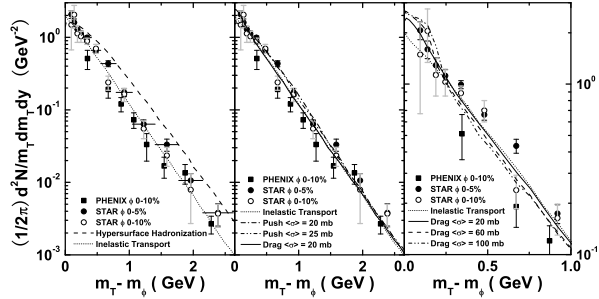}
 \caption{Model estimations for $m_T$ distribution of $\phi(1020)$
    comparing with the data from STAR Collaboration \cite{sign:exp_STAR_phi}
    and the PHENIX Collaboration \cite{sign:exp_PHENIX_phi}.
    (a) Comparison between hypersurface hadronization with transport emit distributions \cite{sign:transMod} and inelastic transport simulation.
    (b) Demonstration of the push and drag effects by elastic collisions.
    (c) Demonstration of drag effects at different mean elastic cross sections.
     Subscripts for the mean elastic cross sections are removed for clarity.
     }
 \label{fig:phi}
\end{figure}

\begin{figure}[h]
    \centering
    \includegraphics[width=0.4\textwidth]{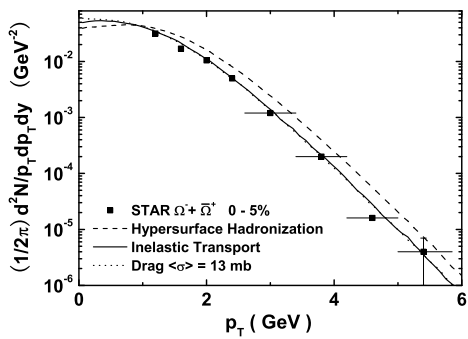}
    \caption{Model estimations for $p_T$ distribution of $\Omega^-+{\bar{\Omega}}^+$
    comparing with the data from STAR Collaboration \cite{sign:exp_STAR_Strange}.
    }
    \label{fig:Omega}
\end{figure}
Inelastic simulations are applied with elastic cross sections
omitted ($Q=0$). For the consistency with \cite{sign:transMod}, no
parameter is changed. There is no new parameter either. The
results for $\phi$ and $\Omega$ are shown in Fig. \ref{fig:phi}a
and Fig. \ref{fig:Omega}. The simulated results look much better
than the original estimations from hypersurface emissions
\cite{sign:transMod}. The productions in very low $p_T$ region are
enhanced and the collective flows are reduced. The perfect
$\Omega$ spectrum shows the elastic collisions for $\Omega$ could
be neglected as they may not participate in the collective
motions. As the enhancement in very low $p_T$ region for $\phi$ is
still smaller than expected, elastic collisions are required in
our considerations.

\section {Elastic Collisions} %
\label{sec:elastic}
The elastic process can be divided to two effects approximately.
The acceleration pushed by the pressure gradient and the
deceleration dragged by the relative motion. Different from the
inelastic cross sections, which is nearly stable via energy scales
and parton species, the elastic cross sections vary from
$10^1\;mb$ to $10^3\;mb$ \cite{sign:PDG2006}, which makes it
difficult to obtain a strict calculation with limited information
of parton-hadron or hadron-hadron elastic differential cross
sections, especially in the condition that only a few selected
hadrons are simulated.

Heavy hadrons are discussed here to present a rough estimation on
elastic process. As they are heavy, the changes of their momenta
after each collision and the thermalization are supposed to be
small. Thus, the push and drag effects can be estimated by
integrating all of the collisions or the probabilities of them.

The push effect can be described in the local rest frame of the
medium,
\begin{equation}
 \left.\frac{d\mathbf{p}_i}{dt}\right|_{push} =
 -\langle{V}\rangle_i\gamma_i^{-1}\nabla{P},
 \label{eq:push}
\end{equation}
where $\langle{V}\rangle_i$ is the mean hadron volume of elastic
collisions. As detailed differential cross sections are not clear,
it is estimated by a rigid ball approximation,
\[
\langle{V}\rangle_i=\frac{4\pi}{3}\left(\frac{\langle\sigma\rangle_{ip}}{\pi}\right)^{\frac{3}{2}}.
\]

On the other hand, the drag effect for a flat plate moving in
normal direction is presented as,
\[
 \left.\frac{d\mathbf{p}_i}{\Delta{S}dt}\right|_{drag}\!\!\!\!\! =
     \sum_j\int{\!\!d^3\mathbf{q_j}f_j(q_j)
    |\mathbf{v}_{j}-\mathbf{v}_{i}|_{\parallel}\triangle{\mathbf{p}_{ij}}(\mathbf{p}_i,\mathbf{q}_j)
    },
\]
where $\triangle{\mathbf{p}_{ij}}(\mathbf{p}_i,\mathbf{q}_j)$ is
the momentum shift after each collision. The integral could be
simplified by assuming the medium particles to be massless and
fixing some of the momentum symbols to a kind of average
$\langle{p_j}\rangle$. Thus,
\begin{equation}
 \left.\frac{d\mathbf{p}_i}{dt}\right|_{drag} =
 -2w_iP(3+v_i^2)\langle\sigma\rangle_{id}\mathbf{v}_i,
 \label{eq:drag}
\end{equation}
where,
\begin{equation}
 w_i = c_f\frac{E_i(E_i + \langle{p_j}\rangle )}{[E_i(1+v_i)+\langle{p_j}\rangle][E_i(1-v_i)+\langle{p_j}\rangle]},
\end{equation}
with $\langle{p_j}\rangle\approx 500\;MeV$ for medium particles
near $T_c \approx 166\;MeV$. $c_f$ is a correction function for
the "shape" of the hadrons.

It should be noted, as the elastic cross sections vary in a large
range for different energy scales, different averages of elastic
cross sections $\langle\sigma\rangle_{ip}$ and
$\langle\sigma\rangle_{id}$ will be reached by equations
(\ref{eq:push}) and (\ref{eq:drag}). The relation between them is
not clear. As Eq (\ref{eq:push}) affects slow hadrons more
intensively and Eq (\ref{eq:drag}) affects fast hadrons more
significantly, $\langle\sigma\rangle_{ip}$ might be slightly
larger than $\langle\sigma\rangle_{id}$. To make the value
$\langle\sigma\rangle_{id}$ similar to
$\langle\sigma\rangle_{ip}$, we estimate $c_f \sim 0.25$ as a
rough approximation.

The elastic corrections are only applied in the medium before it
decouples. Equations (\ref{eq:push}) and (\ref{eq:drag}) are
inserted to the simulation separately. The results are troublesome
when the push and drag effects are inserted together. They disturb
each other in their invalid regions and the problem is not
resolved so far. As shown in Fig. \ref{fig:phi}b, $\phi$
production at very low $p_T$ is enhanced by the drag effect of
preventing hadrons from moving inside. The enhancement
\cite{sign:exp_STAR_phi} could be well satisfied by fitting
$\langle\sigma\rangle_{id} = 20\;mb$. In higher $p_T$ regions,
push effect will effectively accelerate the hadrons, like the
hydrodynamics does, to participate in the collective motions until
the medium decouples, when $\langle\sigma\rangle_{ip}=20\sim 25
\;mb$. They don't work well in opposite regions. The effects of
different $\langle\sigma\rangle_{id}$ are shown in Fig.
\ref{fig:phi}c. The result for $\Omega$ implies the elastic cross
section for $\Omega$ is no more than $20\;mb$.

\begin{figure}[h]
    \centering
    \includegraphics[width=0.4\textwidth]{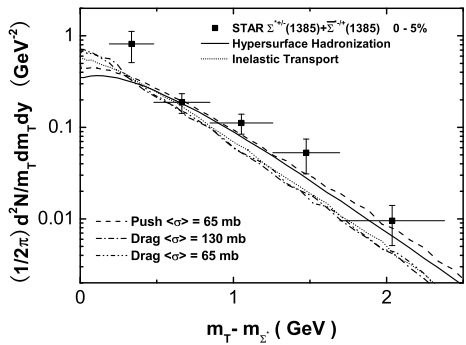}
    \caption{Model estimations for inclusive $m_T$ distribution of $\Sigma^{*\pm}(1385) + \overline{\Sigma}^{*\mp}(1385)$
    comparing with the data from STAR Collaboration
    \cite{sign:exp_STAR_Sig1385}. $\langle\sigma\rangle$ for
    mesons is $(\frac{2}{3})^{2/3}$ of the values listed.
    }
    \label{fig:Sig1385}
\end{figure}

As the momentum of each surviving daughter hadron may be shifted
by elastic collisions, it is hard in current simulations to
determine the efficiency of the reconstruction and the changed
momenta of the reconstructed hadrons, which are different from
those when they decay, until all of the related hadrons including
backgrounds are simulated to commit a real reconstruction. Further
more the elastic cross sections may be much different for hadrons
and their decay products. Thus, The spectra for some strong decay
hadrons, can not be simulated strictly. The spectrum of
$\Sigma^{*}(1385)$ is not satisfied at very low $p_T$ so far, as
shown in Fig. \ref{fig:Sig1385}. $\Xi^{*}(1530)$ has the same
problem but looks more serious. Other possibilities may be the
lack of information for some of heavier baryons which may decay to
$\Sigma^{*}(1385)$ or $\Xi^{*}(1530)$, and inelastic hadron
re-scattering.

\begin{figure}[h]
    \centering
    \includegraphics[width=0.4\textwidth]{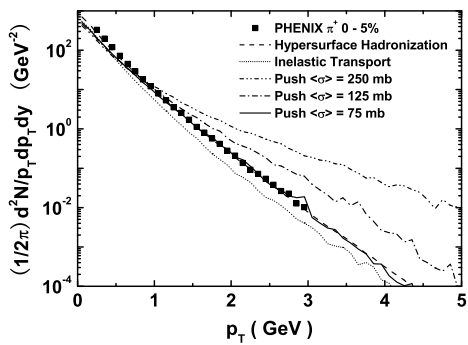}
    \caption{Model estimations for inclusive $p_T$ distribution of $\pi^+$
    comparing with the data from PHENIX Collaboration \cite{sign:exp_PHENIX_ppik}.
    $\langle\sigma\rangle$ for
    baryons is $(\frac{3}{2})^{2/3}$ of the values listed.
    }
    \label{fig:pion}
\end{figure}

Although the equations (\ref{eq:push}) and (\ref{eq:drag}) are not
suitable for light hadrons, we still applied them to test the push
effect for pions. When $\langle\sigma\rangle_{ip}\sim 75\;mb$, the
inclusive pion distribution \cite{sign:exp_PHENIX_ppik} can be
reproduced, as shown in Fig. \ref{fig:pion}. For direct pions
without any decay contributions,
$\langle\sigma\rangle_{ip}\sim{250}\;mb$. Local thermalization as
a Boltzmann distribution was tried for strongly interacting
regions for all the simulations and no significant difference or
better results were found.

\section {Summary} %
\label{sec:summary}

Beyond the approximation of hypersurface hadronization developed
from the transport model \cite{sign:transMod}, we have applied
transport simulations for heavy hadrons to reproduce the
significant enhancement in very low $p_T$ region in some of
transverse momentum spectra. Although the method is neither strict
nor complete, and the values of mean elastic cross sections are
just qualitative, the reason for the enhancement and therefore the
different slopes in the spectra may be successfully explained by
the push and drag effects. The simulation shows that the elastic
cross sections of $\phi$ and $\Omega$ are much smaller than those
of other hadrons. The simulation can also explain the reason why
the hypersurface hadronization works well for light hadrons by the
push effect.

\vspace{2pt}
{\bf Acknowledgements:} We thank Professor Pengfei Zhuang for
useful discussions and Xianglei Zhu for and numerical supports.
This work is supported in part by the grants No. NSFC10547001 and
10425810.


\end{document}